\documentclass[
  aps,
  pra,
  reprint
]{revtex4-1}

\usepackage[english]{babel} 						
\usepackage[latin1]{inputenc} 					
\usepackage{times} 										
\usepackage{graphicx}
\usepackage{amsmath}										
\usepackage{amssymb}							
\usepackage{bbm}

\begin{document}


\newcommand{\ket}[1]{|#1\rangle}
\newcommand{\bra}[1]{\langle #1|}
\newcommand{\ketpsi}{\ket{\psi}}
\newcommand{\brapsi}{\bra{\psi}}
\newcommand{\ketx}{\ket{x}}
\newcommand{\brax}{\bra{x}}
\newcommand{\drho}{\hat{\rho}}
\newcommand{\po}{\hat{p}}
\newcommand{\xo}{\hat{x}}
\newcommand{\Chi}{{\text{\Large$\chi$}}}
\newcommand{\nm}[1]{\|#1\|}
\newcommand{\tr}{\text{tr}}
\numberwithin{equation}{section}
\renewcommand{\theequation}{\arabic{section}.\arabic{equation}}


\title{Optical homodyne tomography with polynomial series expansion}

\author{Hugo Benichi}
\email{hugo.benichi@m4x.org}
\affiliation{Department of Applied Physics, The University of Tokyo, Tokyo, Japan}
\author{Akira Furusawa}
\affiliation{Department of Applied Physics, The University of Tokyo, Tokyo, Japan}

\date{\today}

\begin{abstract}
We present and demonstrate a method for optical homodyne tomography based on the inverse Radon transform.
Different from the usual filtered back-projection algorithm, this method uses an appropriate polynomial series to expand the Wigner function and the marginal distribution and discretize Fourier space.
We show that this technique solves most technical difficulties encountered with kernel deconvolution based methods and reconstructs overall better and smoother Wigner functions. 
We also give estimators of the reconstruction errors for both methods and show improvement in noise handling properties and resilience to statistical errors.
\end{abstract}

\maketitle


\section{Introduction}

In quantum mechanics it is not possible to directly observe a quantum state $\ketpsi$.
In order to obtain full knowledge about $\ketpsi$ it is necessary to accumulate measurement statistics of observables, such as position $\xo$ or momentum $\po$, on many different bases.
In quantum optics, this statistical measurement can be achieved by angle resolved homodyne measurement of the operator $\xo_\theta = \xo\cos\theta + \po\sin\theta$ to acquire statistics of the squared modulus of the wave function $|\langle x_\theta\ketpsi|^2$.
Instead of the quantum state $\ketpsi$, one is rather usually interested in reconstructing the more general density matrix $\drho$ of the system.
Fully equivalent to $\drho$, it is also possible to reconstruct the Wigner function $W(q,p)$ from $|\langle x_\theta\ketpsi|^2$.
However, the reconstruction of $\drho$ or $W(q,p)$ is not immediate and requires the reconstruction of the complex phase of the quantum system from the many angle resolved measurements.
With the measurement of $|\langle x_\theta\ketpsi|^2$, these two operations together are referred to as quantum homodyne tomography or optical homodyne tomography \cite{Smithey93}.

While some tomography algorithms reconstruct the former density matrix, others rather reconstruct the latter Wigner function.
Independently, tomography algorithms can be roughly classified into two species.
Historically the first to be proposed and used for optical homodyne tomography, linear methods exploit and inverse the linear relationship between the experimentally measurable quantity $|\langle x_\theta\ketpsi|$ on one hand and $\drho$ or $W(q,p)$ on the other hand.
Among them, the filtered back-projection algorithm \cite{Vogel89,Smithey93} based on the inverse Radon transform \cite{Radon17} is the most commonly used.
Similar in nature, there also exist methods based on quantum state sampling of individual components of the density matrix $\drho$ with sample functions \cite{Dariono95,Leonhardt96}.
The linear methods, however, suffer in general from technical difficulties associated with the numerical deconvolution necessary to perform the linear inversion of the Radon transform (see Sec. II for details).
In addition, they usually do not guarantee the physicality of the reconstructed state, the positivity of $\drho$.
Finally they perform weakly against statistical noise and show numerical instabilities for higher frequency components and fine details of the reconstructed objects.
Variational methods, such as the maximum entropy \cite{Drobny02} and maximum likelihood \cite{Hradil97} algorithms, were latter applied to optical homodyne tomography to address these problems.
These methods can be designed to enforce the physicality of the reconstructed state and are usually more resilient to statistical errors.
Since the reconstructed states are not defined constructively, an approximation procedure, typically iterative, is used to achieve the reconstruction in practice \cite{Lvovsky04}.

Notice that in theory it is actually possible to bypass these numerical reconstructions and directly observe the Wigner function $W(q,p)$ with repeated measures of the parity operator $\hat{P} = e^{i\pi\hat{n}}$ where $\hat{n}$ is the number operator \cite{Moya93}.
This measurement technique uses the link between the Wigner function value at point $(q,p)$ and the expectation value of $\hat{P}$ for the displaced density matrix $\drho$
\begin{equation}  
W(q,p) = \frac{2}{\pi}
\text{tr}\left[ \hat{D}(-\alpha) \drho \hat{D}(\alpha) e^{i\pi\hat{n}}\right],
\end{equation}
where $\hat{D}$ is the displacement operator and $\alpha = (q+ip)/\sqrt{2}$.
A close tomography technique has been experimentally demonstrated in coupled systems of atoms and light \cite{Deleglise08}.
Unfortunately, a parity detector is a highly non-linear detector which can only be partially implemented for light beams with time-multiplexing and single photon detectors.
Therefore with current state-of-the-art technologies in quantum optics, it is not possible to rely on count statistics alone for quantum state tomography and one has to use optical homodyne tomography based on Gaussian measurements.

While the linear methods look inferior to the variational methods, most of their associated problems are only technical in nature and can in principle be solved.
In this paper we show that is it possible to use a linear reconstruction algorithm with better resilience to noise and better physical properties overall than the usual filtered back-projection method.
The success of this approach lies in a systematic expansion of both the Wigner function $W(q,p)$ and the marginal distribution $p(x,\theta)$ in polar coordinates.
This circular harmonic expansion technique has been applied in the past to other problems where the Radon transform plays a role in tomography \cite{Hawkins86,Rouze06}, and here we adapt it to the quantum framework of optical homodyne tomography.
In Sec. II we first review the basics of the inverse Radon transform and the usual filtered back-projection algorithms for optical homodyne tomography.
In Sec. III we introduce the expansion method:
we first conduct a spectral analysis of the angular components of $p(x,\theta)$ and $W(q,p)$;
from this analysis we argue that a polynomial approximation is an efficient way to expand the radial components.
In Sec. IV we give details about the implementation of the algorithm and also provide an estimator of the reconstruction errors.
Using our estimator we study the performances relatively to the filtered back-projection algorithm on simulated and experimental data sets.
We complete this comparison with numerical studies of the distance between target and reconstructed quantum states.


\section{Filtered back-projection}

In 1917, Radon introduces the integral transform $\mathcal{R}$ of two-dimensional functions integrated along straight lines and provides the formula for the inverse transform $\mathcal{R}^{-1}$ \cite{Radon17}.
Today the Radon and inverse Radon transforms are ubiquitous in tomography and find applications in many different area of science. 
The Radon transform is as well applicable to optical homodyne tomography.
First we recall the definition of the observable operator $\xo_\theta$ of an homodyne measurement,
\begin{equation}
\label{2_homodyne}
\xo_\theta = \hat{U}_\theta^\dagger \xo \hat{U}_\theta = \xo\cos\theta + \po\sin\theta,
\end{equation}
where $\hat{U}_\theta$ is the rotation operator in phase space, or phase-shifting operator.
The marginal distribution of the homodyne current $p(x,\theta)$ is then distributed according to the squared modulus of the wave function
\begin{equation}
\label{2_marginal}
p(x,\theta)
= | \langle x_\theta \ketpsi |^2
=  \bra{x} \hat{U}_\theta \ketpsi \brapsi \hat{U}^\dagger_\theta \ket{x},
\end{equation}
where $ |x_\theta \rangle$ is the eigenvector of $\xo_\theta$.
The Radon transform $\mathcal{R}$ links the Wigner function $W(q,p)$ of the quantum state $\ketpsi$ and $p(x,\theta)$ the marginal distribution of the homodyne current with a projection of $W(q,p)$ on a particular angle of observation $\theta$ \cite{Leonhardt}
\begin{eqnarray}
\label{2_radon_trans}
p(x,\theta) & = & \mathcal{R}\left(W\right) \nonumber\\
& = & \iint_{\mathbbm{R}^2}  W(q,p) \delta( x - q\cos\theta-p\sin\theta) dq dp \nonumber\\
& = & \int_{-\infty}^{+\infty} W(x\cos\theta-p\sin\theta,x\sin\theta+p\cos\theta) dp. \qquad
\nonumber\\&&
\end{eqnarray}
In his original paper, Radon mathematically inverses his transform with the back-projection $\mathcal{B}$ of the derivative of the Hilbert transform $\mathcal{H}$ of $p(x,\theta)$
\begin{eqnarray}
\label{2_backproj_hilbert}
W(q,p) = \frac{1}{2\pi}\mathcal{B}\left(\frac{\partial}{\partial y} \mathcal{H}(p(x,\theta))(y). \right),
\end{eqnarray}
where the back-projection operator $\mathcal{B}$ of a function $f(x,\theta)$ is the function $F(q,p)$ defined by
\begin{equation}
F(q,p) = \int_0^{\pi} f(q\cos\theta+p\sin\theta,\theta) d\theta.
\end{equation}
Expanding Eq. \eqref{2_backproj_hilbert} we obtain the inversion formula
\begin{equation}
\label{2_inv_radon_trans}
W(q,p) = -\frac{\mathcal{P}}{2\pi^2} \int_0^\pi \int_{-\infty}^{+\infty}
\frac{p(x,\theta)}{(q\cos\theta+p\sin\theta-x)^2} dx d\theta,
\end{equation}
where $\mathcal{P}$ is the principal-value operator.
Although exact, this expression is nevertheless unusable with experimental data as the algebraic expression of $p(x,\theta)$ is unknown.
\begin{figure}[!ht]
  \includegraphics[height=\columnwidth,angle=-90]{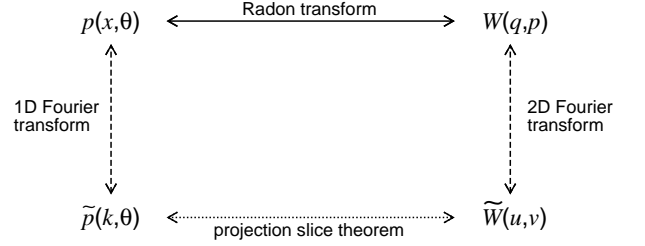}
  \caption{\label{fig_transform_map}Different transforms for different paths from $p$ to $W$.}
\end{figure}
However, the projection-slice theorem or Fourier slice theorem \cite{Bracewell56} gives another reverse path from $p(x,\theta)$ to $W(q,p)$ to work around the difficulties of the principal-value operator (see Fig.\ref{fig_transform_map}).
If $\tilde{p}(k,\theta)$ and $\tilde{W}(u,v)$ are, respectively, the one-dimensional and two-dimensional Fourier transforms of $p(x,\theta)$ and $W(q,p)$, then the projection-slice theorem states that
\begin{equation}
\label{2_projection_slice}
\tilde{p}(k,\theta) = \tilde{W}(k\cos\theta,k\sin\theta).
\end{equation}
Simply computing the Fourier transform $\tilde{p}(k,\theta)$ from the measured data would seem like the most efficient way to obtain $W(q,p)$ after a second inverse Fourier transform, but Eq. \eqref{2_projection_slice} shows that it is necessary to interpolate $\tilde{W}(u,v)$ in Fourier space, which leads to significant numerical difficulties \cite{Stark81}.
To avoid this interpolation Eq. \eqref{2_projection_slice} can be used to replace $\tilde{W}(u,v)$ in the inverse Fourier transform of $W(q,p)$ to obtain the inversion formula,
\begin{equation}
\label{2_back_filtered_proj}
W(q,p) = \frac{1}{2\pi} \int_{0}^\pi \int_{-\infty}^{+\infty} 
p(x,\theta) K(q\cos\theta+p\sin\theta-x) dx d\theta.
\end{equation}
Here, the marginal distribution is convoluted with an integration kernel $K(x)$ and then back-projected into phase space, where $K(x)$ is defined as the inverse Fourier transform of $|k|$
\begin{equation}
\label{2_kernel}
K(x) = \frac{1}{2\pi} \int_{-\infty}^{+\infty} |k| e^{ik x} dk.
\end{equation}
To use Eq. \eqref{2_back_filtered_proj} in practice it is necessary to regularize $K(x)$ and replace it with some numerical approximation.
This is possible with the use of a window function $g(k)$ such that the integral,
\begin{equation}
\label{2_windowed_kernel}
\frac{1}{2\pi}\int_{-\infty}^{+\infty} g(k) |k| e^{ik x} dk,
\end{equation}
converges.
\begin{figure}[!ht]
  \includegraphics[height=\columnwidth,angle=-90]{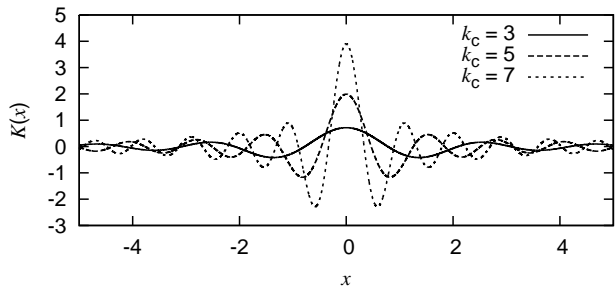}
  \caption{\label{fig_kernel}Regularized integration kernel $K(x)$ for different values of $k_c$.}
\end{figure}
The most common way to regularize Eq. \eqref{2_kernel} is to choose $g(k) = \mathbbm{1}_{[-k_c,+k_c]}(k)$ and introduce a hard frequency cutoff parameter $k_c$ so that
\begin{equation}
K(x) \approx \frac{1}{\pi x^2} \left( \cos(k_cx) + k_cx \sin(k_cx) -1 \right).
\end{equation}

In practice, the choice of $k_c$ affects how much high frequency components of the Wigner function will get reconstructed.
If $k_c$ is set too low the convolution in Eq .\eqref{2_back_filtered_proj} will filter out the fine physical details of the Wigner function.
If $k_c$ is set too high, the convolution will introduce unphysical high frequency noise from the statistical errors in the measurement of $p(x,\theta)$.
Figure \ref{fig_kernel} shows the integration kernel $K(x)$ for different high frequency sensitivities.
Choosing the right value of $k_c$ is a trade off between these two regimes.
From Eqs. \eqref{2_projection_slice} and \eqref{2_back_filtered_proj} it is also possible to insert other filter functions at different steps of the inversion to obtain modified algorithms with enhanced and more selective noise filtering properties.
In any case the numerical implementation of Eq. \eqref{2_back_filtered_proj} will rely on deconvolution of the marginal distribution, an operation very sensitive to statistical noise.


\section{Harmonic series expansion}

To numerically perform optical homodyne tomography, it is necessary at some point to apply an approximation procedure from the infinite dimensional space which features the unknown physical state to a finite dimensional space used to describe the reconstructed state.
In the filtered back-projection algorithm, the discretization is achieved by direct evaluation of $W(x_i,p_i)$ on the set of points $\{(x_i,p_i)\}_i$ chosen to probe the phase space.
Rather than this point-by-point reconstruction, a discretization of another space should help to solve the numerical issues encountered in Sec. II.
Since we are dealing with objects behaving like probability distributions, the statistical moments of $p(x,\theta)$ and $W(q,p)$ might be a solution to the problem.
In Ref.\cite{Ourjoumtsev06}, Ourjoumtsev {\it et al.} describes such a technique where they parametrize the Wigner function of a photon subtracted squeezed vacuum with the second and fourth moments of the marginal distribution $p(x,\theta)$.
Generalizing this approach for any quantum state to higher order moments requires the use of the moment generating function $\left\langle e^{\lambda x}\right\rangle $, where $\langle x\rangle$ is the expectation value of $x$ with regards to $p(x,\theta)$.
Superior to the moment generating function the characteristic function $\left\langle e^{i\lambda x}\right\rangle $ only needs the mean and variance to be defined to exist.
This and the projection-slice theorem of Eq. \eqref{2_projection_slice} hint that Fourier space is a good candidate for an efficient discretization.

We decompose our discretization procedure in two steps: (1) an angular harmonic decomposition with Fourier series; (2) a polynomial series expansion of the radial components.
We express $W(q,p)$ in radial coordinates $(r,\phi)$ and notice that $W(r,\phi+2\pi) = W(r,\phi)$.
Therefore we write the radial part of $W(r,\phi)$ in terms of a Fourier series and we define the set of radial functions, or angular harmonic components $\{w_n(r)\}_n$ by
\begin{equation}
w_n(r) = \frac{1}{2\pi}\int_{-\pi}^{+\pi} W(r,\phi) e^{-in\phi} d\phi,
\end{equation}
which allows us to write $W(r,\phi)$,
\begin{equation}
\label{3_wigner_angular_decomposition}
W(r,\phi) = \sum_{n=-\infty}^\infty w_n(r)e^{in\phi}, 
\end{equation}
with the symmetry relation $w_n(r) = w_{-n}^*(r)$. The 2D Fourier transform $\tilde{W}(u,v)$ of $W(q,p)$ is written in radial coordinates,
\begin{equation}
\tilde{W}(k,\theta) = \int_0^{+\infty}\int_{-\pi}^{+\pi} W(r,\phi) e^{-irk\cos(\theta-\phi)} rdrd\phi,
\end{equation}
with the change of variables $(u,v)\rightarrow(k,\theta)$.
$\tilde{W}(u,v)$ is related to the Weyl function $\Chi(u,v) = \text{tr}(\drho e^{-iv\hat{q}+iu\po})$ by a simple $\pi/2$ rotation,
\begin{eqnarray}
\tilde{W}(u,v) &=& \Chi(-v,u), \\
\tilde{W}(k,\theta) &=& \Chi(k,\theta+\frac{\pi}{2}).
\end{eqnarray}
We can easily write $\tilde{W}$ in polar coordinates in terms of the angular harmonic components $w_n(r)$ of $W(r,\phi)$,
\begin{eqnarray}
\label{3_chi_fourier_transform}
\tilde{W}(k,\theta) &=& 
\sum_{n=-\infty}^\infty \int_0^{+\infty} w_n(r) rdr \nonumber\\
&\times& \int_{-\pi}^{+\pi} e^{-ikr\cos(\theta-\phi)+in\phi} d\phi.
\end{eqnarray}
With a Jacobi-Anger expansion of $e^{iz\cos\phi}$ using Bessel functions $J_n$,
\begin{equation}
e^{iz\cos\phi} = \sum_{n=-\infty}^{\infty} i^n J_n(z) e^{in\phi},
\end{equation}
it is possible to conduct the angular integration in Eq. \eqref{3_chi_fourier_transform} to obtain the expression,
\begin{equation}
\tilde{W}(k,\theta) 
= 2\pi\sum_{n=-\infty}^\infty (-i)^n e^{in\theta} \int_0^\infty w_n(r) J_n(kr) rdr.
\end{equation} 
Notice that $\int_0^\infty w_n(r) J_n(kr) rdr$ is the $n^{\text{th}}$ order Hankel transform of $w_n(r)$.

In the same fashion, since $p(x,\theta+2\pi) = p(r,\theta)$ we decompose the marginal distribution as
\begin{equation}
p_\theta(x) = \sum_{n=-\infty}^\infty c_n(x)e^{in\theta},
\end{equation}
with the sets of radial functions $c_n(x)$ defined by
\begin{equation}
c_n(x) = \frac{1}{2\pi}\int_{-\pi}^{+\pi} p(x,\theta) e^{-in\theta} d\theta.
\end{equation}
Using the projection-slice theorem of Eq. \eqref{2_projection_slice} and the orthogonality of $e^{in\theta}$ on $[-\pi,+\pi]$ we are able to write for every angular harmonic order $n$,
\begin{equation}
\label{3_fourier_equal_hankel}
\frac{i^n}{2\pi} \int_{-\infty}^{+\infty} c_n(x)e^{-ikx} dx = \int_0^\infty w_n(r) J_n(kr) rdr.
\end{equation}
We have obtained a relation between, on one side the Fourier transform of the angular harmonics of $p(x,\theta)$, and on the other side, the Hankel transform of the angular harmonics of $W(r,\phi)$.
If we inverse the Hankel transform with the orthogonality relation, or closure relation of Bessel functions,
\begin{equation}
\int_0^{\infty}kdkJ_n(kr)J_n(kr') = \frac{1}{r}\delta(r-r'),
\end{equation}
we finally obtain 
\begin{equation}
\label{3_angular_inversion}
w_n(r) = \frac{i^n}{2\pi}\int_0^\infty J_n(kr) kdk \int_{-\infty}^{+\infty} c_n(x)e^{-ikx} dx.
\end{equation}

At that point it would be natural to convey some radial decomposition of $w_n(r)$ and $c_n(x)$ . However, there is no simple way to achieve this.
Looking at Eq. \eqref{3_angular_inversion}, we notice that the Fourier transform of $kJ_n(k)$, or at least $J_n(k)$, should be involved in the process.
The latter is written in terms of the Chebysheff's polynomials of the first kind $T_n$
\begin{equation}
\label{3_bessel_fourier_transform}
\int_{-\infty}^{+\infty} J_n(k) e^{-ikx} dk
= \frac{2(-i)^n}{\sqrt{1-x^2}} T_n(x) \mathbbm{1}_{[-1,+1]}(x).
\end{equation}
Equation \eqref{3_bessel_fourier_transform} hints at the use of the polynomial series to achieve this radial decomposition.
It is safe to assume for applications that the Wigner function will only take nonzero values from the origin up to a certain limit $L\geq r$. 
Since we are carrying the decomposition in polar coordinates what we are looking after is a polynomial family which is orthogonal on a disk of radius $L$.
There are of course infinitely many such families but one which proves to be particularly adequate to the task is the set of Zernike polynomials $Z_s^n(r,\varphi) = R_s^n(r) e^{in\varphi}$ originally introduced for the study of optical aberrations in lenses and other circular optical systems \cite{BornWolf}.
The polynomials are defined for $s\geq |n| \geq 0$ and $s-|n|$ even.
While the angular part gives straightforward orthogonality and fits with our previous approach using Fourier series, the radial components $R_s^{\pm n}$ defined for $t = |n|\geq 0$ by
\begin{equation}
\label{3_zernike_radial}
R_s^{\pm n}(r) = \sum_{k=0}^{(s-t)/2} (-1)^k
\frac{(s-k)!}{k!\left(\frac{s+t}{2}-k\right)!\left(\frac{s-t}{2}-k\right)!} r^{s-2k},
\end{equation}
are orthogonal on $[0,1]$ with respect to the weight function $r$ for all positive and negative orders $n$,
\begin{equation}
\label{3_zernike_orthogonals}
\int_0^1 R_s^n(r) R_{s'}^n(r) r dr = \frac{1}{2(s+1)} \delta_s^{s'}.
\end{equation}
Furthermore it turns out that the Radon transform of Zernike polynomials happens to have the simple expression,
\begin{equation}
\label{3_zernike_radon}
\mathcal{R}\left(R_s^n(r)e^{in\phi}\right) = 
\frac{2}{s+1}\sqrt{1-x^2} U_s(x) e^{in\theta},
\end{equation}
where $U_s(x)$ are the Chebysheff's polynomials of the second kind \cite{Deans07,Zeitler74} (see also the last paragraph of this section for a proof).
The critical aspect for tomography lies in the fact that $U_s(x)$ is again an orthogonal polynomial family on $[-1,1]$ with respect to the weight function $\sqrt{1-x^2}$.
In other words by finding a family of orthogonal polynomials whose Radon transform element by element is yet another family of orthogonal polynomials, we have in some sense diagonalized the Radon transform.
The inverse Radon transform can also be exactly calculated and any technical difficulties associated with kernel functions or regularization immediately vanish.

With the use of Eq. \eqref{3_zernike_orthogonals} we are eventually able to expand the angular harmonic functions $w_n(r)$ on the $n^{\text{th}}$ order radial polynomials $R_s^n(r)$,
\begin{equation} 
\label{3_radial_decomposition}
w_n(r) = \sum_{s=0}^\infty w_n^s R_s^n(r).
\end{equation}
Given that $R_s^n(r)$ is non zero only when $s\geq |n| \geq 0$ and $s-|n|$ is even, we introduce the change of variable $s \rightarrow |n|+2m$, re-index the sequence $w_n^s$ and rewrite Eq. \eqref{3_radial_decomposition}
\begin{equation} 
\label{3_radial_decomposition_bis}
w_n(r) = \sum_{m=0}^\infty w_n^m R_{|n|+2m}^n(r).
\end{equation}
Putting Eqs. \eqref{3_radial_decomposition_bis} and \eqref{3_wigner_angular_decomposition} together we obtain the complete expansion of $W(r,\phi)$ inside the unit disk $D(0,1)$,
\begin{equation}
\label{3_full_wigner_expansion}
W(r,\phi) = \sum_{n=-\infty}^\infty \sum_{m=0}^\infty w_n^m R_{|n|+2m}^{|n|}(r) e^{in\phi}. 
\end{equation}
Notice from Eq. \eqref{3_zernike_radial} that $R_s^{+n}(r) = R_s^{-n}(r)$ which justifies the use of $R_{|n|+2m}^{|n|}$ although $w_n^m$ are in general complex constants.
Applying the relation \eqref{3_zernike_radon} on Eq. \eqref{3_full_wigner_expansion}, $p(x,\theta)$ is also written in terms of the coefficients $w_n^m$ as
\begin{equation}
\label{3_chebysheff_marginal}
p(x,\theta) = \sum_{n=-\infty}^\infty \sum_{m=0}^\infty 
\frac{2w_n^m}{|n|+2m+1} \sqrt{1-x^2} U_{|n|+2m}(x) e^{in\theta}. 
\end{equation}

To justify the use of Zernike polynomials and prove Eq. \eqref{3_zernike_radon}, the relation,
\begin{equation}
\label{2_integral_transform_bessel_zernike}
\int_0^1 R_m^n(r) J_n(rk) rdr = (-1)^{(m-n)/2} \frac{J_{m+1}(k)}{k},
\end{equation}
between Zernike polynomials and Bessel functions \cite{BornWolf} is essential.
If we recall Eq. \eqref{3_fourier_equal_hankel}, replace $w_n(r)$ by its expansion on $R_s^n(r)$ in Eq. \eqref{3_radial_decomposition} and cut the integration from $+\infty$ to unity, we obtain
\begin{equation}
\label{3_Fourrier_decomp_on_bessels}
\sum_{m=0}^\infty w_n^m (-1)^m \frac{J_{|n|+2m+1}(k)}{k}
= \frac{i^{|n|}}{2\pi} \int_{-\infty}^{+\infty} c_n(x)e^{-ikx} dx.
\end{equation}
To finally obtain the complete inversion of $\mathcal{R}$ and the expansion of $c_n(x)$ as in Eq. \eqref{3_chebysheff_marginal}, we only need to inverse the Fourier transform in Eq. \eqref{3_Fourrier_decomp_on_bessels} from the rhs to the lhs and use the Fourier transform of $J_s(k) / k$,
\begin{equation}
\label{2_fourier_transform_of_bessels}
\int_{-\infty}^{+\infty} \frac{J_{s+1}(k)}{k} e^{ikx} dk 
= \frac{2i^s}{s+1} U_{s}(x) \sqrt{1-x^2} \mathbbm{1}_{[-1,+1]}(x),
\end{equation}
to obtain
\begin{equation}
c_n(x) =
\sum_{m=0}^\infty \frac{w_n^m}{|n|+2m+1} U_{|n|+2m}(x) \sqrt{1-x^2} \mathbbm{1}_{[-1,+1]}(x).
\end{equation}
Notice that Eqs. \eqref{2_integral_transform_bessel_zernike} and \eqref{2_fourier_transform_of_bessels} close the link between $U_s(x)$ and $R_m^n(r)$, the first two families of orthogonal functions used in the analysis, and the Bessel functions $J_n(k)$ orthogonal with respect to the weight function $1/k$,
\begin{equation}
\int_0^\infty J_s(k) J_t(k) \frac{dk}{k} = \frac{1}{2s}\delta_s^k
\end{equation}
In summary by identifying three families of orthogonal functions related together by the Radon transform $\mathcal{R}$ and the Fourier transform $\mathcal{F}$, we have been able to find an expansion of the Wigner function $W(q,p)$ that allows to greatly simplify the technical difficulties of tomography with inverse Radon transform.


\section{Reconstruction algorithm}

\subsection{The algorithm}

\begin{figure}[!ht]
\includegraphics[width=0.9\columnwidth]{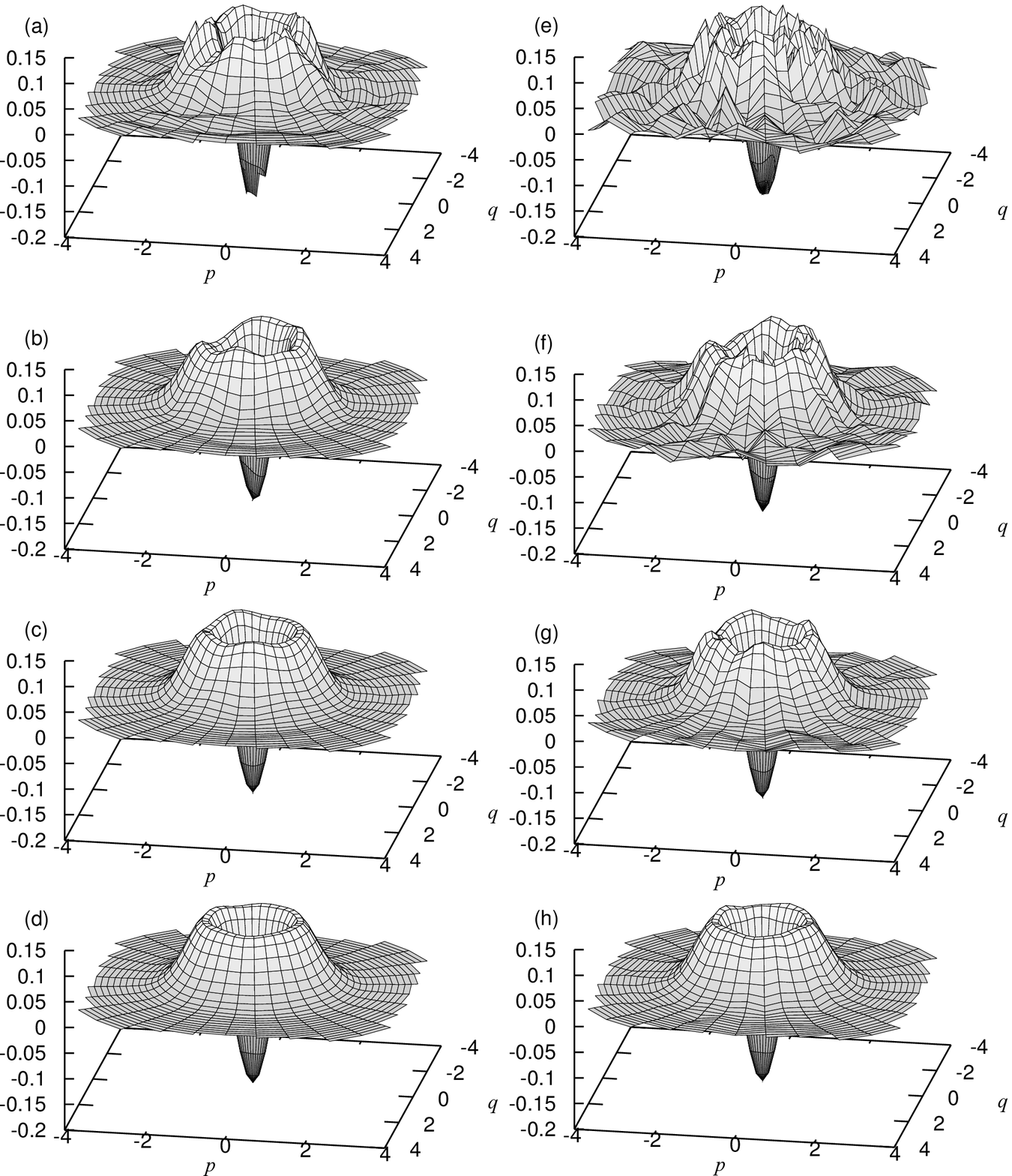}
\caption{\label{fig_photon_comp}
Comparison between polynomial series tomography (left panels: $N = 8, M = 30$) and filtered back-projection tomography (right panels: $k_c = 8.0$) for the state $\rho = 0.8 \ket{1}\bra{1} + 0.2 \ket{0}\bra{0}$. 
(a) $J=5\times10^3$; 
(b) $J=20\times10^3$; 
(c) $J=80\times10^3$; 
(d) $J=320\times10^3$; 
(e) $J=5\times10^3$;
(f) $J=20\times10^3$; 
(g) $J=80\times10^3$; 
(h) $J=320\times10^3$.
All data sets have been synthetically generated with rejection sampling. 
}
\end{figure}

\begin{figure}[!ht]
\includegraphics[width=0.9\columnwidth]{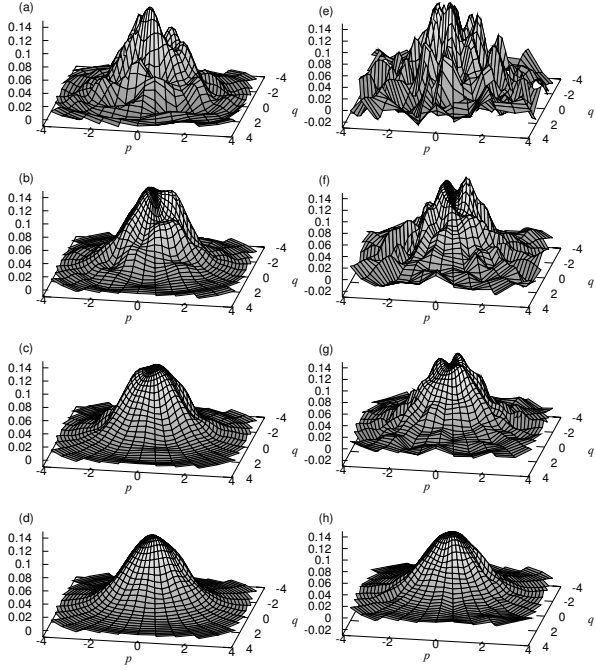}
\caption{\label{fig_thermal_comp}
Comparison between polynomial series tomography (left panels: $N = 8, M = 30$) and filtered back-projection tomography (right panels: $k_c = 8.0$) for a thermal state of mean photon number $\langle\hat{n}\rangle = 1$. 
(a) $J=5\times10^3$; 
(b) $J=20\times10^3$; 
(c) $J=80\times10^3$; 
(d) $J=320\times10^3$; 
(e) $J=5\times10^3$;
(f) $J=20\times10^3$; 
(g) $J=80\times10^3$; 
(h) $J=320\times10^3$.
All data sets have been synthetically generated with rejection sampling. 
}
\end{figure}

The algorithm works in four steps: (1) choosing the size $L$ of the reconstruction disk, (2) evaluating the coefficients $w_n^m$, (3) choosing the cutoffs $N$ and $M$ of the angular and radial series, and (4) calculating $W(r,\phi)$.
Step 1 is necessary for the orthogonal relations given in Sec. II on $[0,1]$ and $[-1,+1]$ to hold.
In practice we have to normalize the marginal distribution $p(x,\theta) \rightarrow p(x/L,\theta)/L $ and the Wigner function $ W(r,\phi) \rightarrow W(r/L,\phi)/L$.
Step 2 is easily conducted by inverting the relation \eqref{3_chebysheff_marginal} with the orthogonal Chebysheff's polynomials $U_{|n|+2m}(x)$,
\begin{eqnarray}
\label{4_coefficients}
w_n^m &=& \frac{|n|+2m+1}{2\pi^2} \int_{-\pi}^{+\pi} d\theta e^{-in\theta} \nonumber\\
      &\times& \int_{-1}^{+1} dx\, \frac{p(x/L,\theta)}{L}  U_{|n|+2m}(x).  
\end{eqnarray}
The recurrence relation, 
\begin{equation}
U_{s+1}(x) = 2x U_s(x) - U_{s-1}(x),
\end{equation}
allows one to efficiently calculate $U_s(x)$ for any $s$ and any $x$ given $U_0(x) = 1$ and $U_1(x) = 2x$.
After obtaining the coefficients $w_n^m$ and choosing cutoff orders $N$ and $M$, the Wigner function $W(r,\phi)$ is then approximated by the partial sums,
\begin{equation}
\label{4_complex_summation}
W'(r,\phi) = \sum_{n=-N}^N \sum_{m=0}^M w_n^m R_{|n|+2m}^{|n|}\left(\frac{r}{L}\right) e^{in\phi} / L,
\end{equation}
Using the symmetry relation $w_{-n}^m = (w_n^m)^*$, we keep the real part of Eq. \eqref{4_complex_summation} and simplify the sum on $n$ to
\begin{eqnarray}
\label{4_real_summation}
W'(r,\phi) &=& \sum_{m=0}^M \sum_{n=0}^N  R_{n+2m}^n\left(\frac{r}{L}\right)/L \nonumber\\
&&\times \left( a_n^m \cos(n\phi) +  b_n^m \sin(n\phi) \right).
\end{eqnarray}
where we have defined $w_n^m = (a_n^m + i b _n^m)/2$ for $n\geq 1$ and $w_0^m = a_0^m$.
Figures \ref{fig_photon_comp} and \ref{fig_thermal_comp} show examples of reconstructed Wigner functions for a mixture of $\ket{0}$ and $\ket{1}$, and a thermal state respectively.
In comparison to filtered back-projection tomography, polynomial series tomography converges faster with fewer numbers of experimental points $J$.
The reconstructed Wigner functions also show less visible artifacts and are overall smoother.
To evaluate efficiently $R_n^m(r)$ we notice that $R_n^n(r) = r^{|n|}$ and then use the recurrence relation \cite{Prata89},
\begin{eqnarray}
R^n_{n+2(m+1)}(r) = \frac{n+2(m+1)}{(m+1)(n+m+1)} \times \left\{\frac{}{}\right. \qquad\qquad\qquad \nonumber\\
\left((n+2m+1)r^2 - \frac{(n+m)^2}{n+2m} - \frac{(m+1)^2}{n+2(m+1)} \right)R^n_{n+2m}(r) \nonumber\\
\left. -\quad m\frac{n+m}{n+2m} R^n_{n+2(m-1)}(r) \quad \right\}. \qquad\qquad
\end{eqnarray}

\begin{figure}[!ht]
\includegraphics[width=0.9\columnwidth]{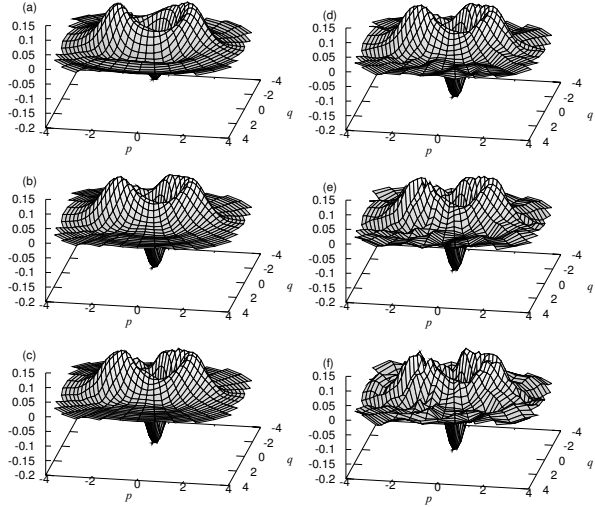}
\caption{\label{fig_input_comp}
Effect of increased radial resolution on the stability of tomography of an experimentally measured photon subtracted squeezed vacuum (same data as in Ref. \cite{Lee11}).
For all panels $J=1\times 10^5$.
(a) Polynomial series tomography, $N=8$, $M=20$;  
(b) $M=30$;
(c) $M=40$;
(d) filtered back-projection tomography, $k_c = 7$;
(e) $k_c = 9$;
(f) $k_c = 11$.
}
\end{figure}

\begin{figure}[!ht]
\includegraphics[width=0.9\columnwidth]{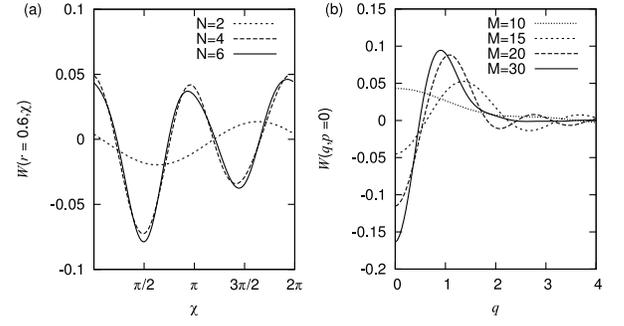}
\caption{\label{fig_series_conv_cuts}
Effect of $N$ and $M$ on the convergence of polynomial series tomography. 
Same experimental data as in Fig. \ref{fig_input_comp}. 
(a) Circular cut at constant $r$ and effect of $N$ for $M=32$, $J=2\times 10^5$.
(b) Radial cut at constant $\phi$ and effect of $M$ for $N=10$, $J=2\times 10^5$.
} 
\end{figure}

\begin{figure}[!ht]
\includegraphics[width=0.7\columnwidth]{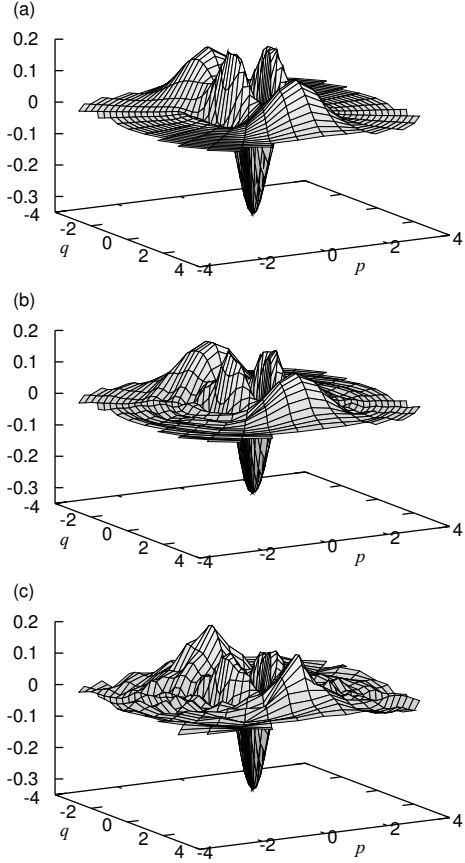}
\caption{\label{fig_cat_tomography} 
Effect of increased radial resolution on the stability of tomography of a Schroedinger's cat states with $\langle \hat{n} \rangle = 3$. 
For all panels $J=4\times 10^4$.
(a) Original Wigner function; 
(b) polynomial series tomography, $N=8$, $M=46$;
(c) filtered back-projection tomography, $k_c = 11$.
}
\end{figure}

In contrast to setting the value of $k_c$, the values of $N$ and $M$ have a real physical meaning.
This is a major advantage of this method compared to the usual filtered back-projection algorithm.
$M$ will decide what will be the highest polynomial order of the radial features of $W$.
Therefore it is equivalent to choosing the maximum photon number of the density matrix diagonal elements.
$N$ will set the resolution of the angular features of $W$, which decides how many off-diagonal components of the density matrix will be reconstructed.
Furthermore it is easy to change $N$ and $M$ after computing the coefficients $w_n^m$.
Figure \ref{fig_input_comp} shows the effect of increasing $M$ on the precision of polynomial series tomography.
In comparison to filtered back-projection tomography when increasing the kernel sensitivity $k_c$, increasing the radial resolution $M$ does not produce artifacts in the Wigner function.
Figure \ref{fig_series_conv_cuts} further shows the effect of increasing $N$ and $M$ on the precision of the tomography reconstruction of experimental data.
While the angular components show quick convergence, the radial components require higher $M$ values to be faithfully reconstructed.
Figure \ref{fig_cat_tomography} illustrates the advantage of polynomial series tomography in radial resolution for quantum states with a higher number of photons.
Both $M$ and $k_c$ where set at values high enough to recover the original Schroedinger's cat state negativity at the origin of phase space.
While the back-filtered projection shows numerical uinstability when $k_c$ is set high, the Wigner function reconstructed by polynomial series tomography is smoother at the equivalent resolution.

Finally the value of $R_{n+2m}^n$ in $r=0$ will be non-zero only for $n=0$, therefore we have the useful formula to evaluate the Wigner function at the origin of phase space,
\begin{equation}
\label{4_a_neg_zernike}
W'(0,0) = \sum_{m=0}^M (-1)^m a_0^m / L,
\end{equation}
which is similar to the formulation of $W(0,0)$ using the diagonal elements of the density matrix.


\subsection{Unbiased error estimator}

To quantitatively compare our algorithm with the usual back-filtered tomography algorithm we give a consistent method to estimate the reconstruction error and obtain confidence intervals when calculating the value of $W(q,p)$.
If $W'$ and $W''$ are the reconstructed value of $W(q,p)$ with Eqs. \eqref{4_real_summation} and \eqref{2_back_filtered_proj} respectively, we call $\sigma^2_{W'}$ and $\sigma^2_{W''}$ the variance of the reconstruction errors assuming they are distributed according to a Gaussian for both algorithms.
We also assume that there are no systematic errors but only statistical errors.
Let's assume an optical homodyne measurement set consists of $J$ experimental points $\{(x_j,\theta_j)\}_j$ independently and identically distributed according to the underlying marginal distribution $p(x,\theta)$. 
To begin with we give an estimator of $\sigma^2_{W''}(q,p)$ for the usual filtered back-projection method using formula \eqref{2_back_filtered_proj}.
To calculate the value of $W$ at point $(q,p)$, $p(x,\theta)$ will be replaced either by a binned histogram made from the data set $\{(x_j,\theta_j)\}_j$, or by a sum of delta functions approximating $p(x,\theta)$
\begin{equation}
\label{4_b_sum_diracs}
p(x,\theta) = \frac{1}{J} \sum_j \delta(x-x_j) \times \delta(\theta-\theta_j).
\end{equation}
In the latter case, the swap of $p(x,\theta)$ for expression \eqref{4_b_sum_diracs} in Eq. \eqref{2_back_filtered_proj} leads to
\begin{equation}
\label{4_b_expectation_kernel}
W''(q,p) = \frac{1}{2\pi J} \sum_{j=1}^J K(q\cos\theta_i+p\sin\theta_i-x_i).
\end{equation}
Since $p(x,\theta)$ is a valid probability distribution $W''(q,p)$ is nothing else than $\langle K(q\cos\theta+p\sin\theta-x) \rangle$ the expectation value of the kernel function.
Therefore Eq. \eqref{4_b_expectation_kernel} can be regarded as a Monte Carlo integral where the expectation value of the kernel function is calculated by randomly sampling $K$ according to the distribution $p(x,\theta)$.
In other words, the optical homodyne tomography with filtered back-projection is in effect an analogical Monte Carlo integration where the homodyne measurement plays the part of the random number generator.
In that familiar case the statistical properties of the reconstruction error are well known.
First of all we are assured of the unbiased convergence of the sum in Eq. \eqref{4_b_expectation_kernel}.
The central limit theorem also states that the error will indeed converge to a Gaussian distribution of zero mean and whose standard deviation $\sigma_{W''}(q,p)$ for $J$ experimental points is
\begin{equation}
\label{4_b_kernelirt_error}
\sigma_{W''}(q,p) = \sigma_{K} / \sqrt{J-1},
\end{equation}
which exhibits a $1/\sqrt{J}$ rate of convergence, and where $\sigma_{K} = \sqrt{\langle K^2\rangle - \langle K\rangle^2} / 2\pi$.
By using the approximations,
\begin{eqnarray}
\label{4_b_kernel_error_estimator}
\langle K\rangle   & \approx & \frac{1}{J} \sum_{j=0}^J K(q\cos\theta_j+p\sin\theta_j-x_j), \qquad \\
\label{4_b_kernel_error_estimator_bis}
\langle K^2\rangle & \approx & \frac{1}{J} \sum_{j=0}^J K^2(q\cos\theta_j+p\sin\theta_j-x_j), \qquad
\end{eqnarray}
we can actually estimate $\sigma_{K}$ in a straightforward way easy to include in the implementation of Eq. \eqref{4_b_expectation_kernel}.

\begin{figure}[!ht]
\includegraphics[width=0.9\columnwidth]{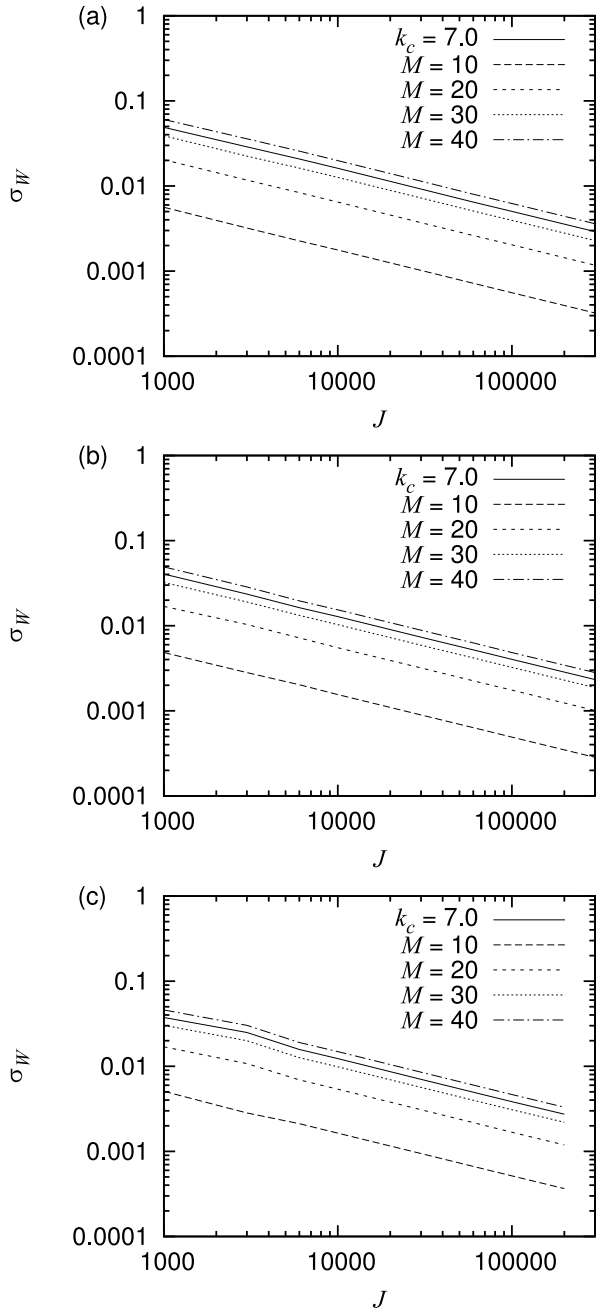}
\caption{\label{fig_estim_err}
Estimation of $\sigma_W(0,0)$ with filtered back-projection tomography (plain line) and polynomial series tomography (dotted lines).
(a) $\rho = 0.8 \ket{1}\bra{1} + 0.2 \ket{0}\bra{0}$;
(b) thermal state with $\langle\hat{n}\rangle = 1$;
(c) photon subtracted squeezed vacuum (same data as in Fig.\ref{fig_input_comp}). 
}
\end{figure}

The same analysis for the coefficients $\{w_n^m\}$ yields the reconstruction sum,
\begin{equation}
\label{4_coefficients_discrete}
w_n^m = \frac{|n|+2m+1}{2\pi^2} \sum_{j=1}^J U_{|n|+2m}(x_j/L) e^{-in\theta_j} / L.
\end{equation}
As previously errors are Gaussian distributed for every coefficient $w_n^m$ with a $1/\sqrt{J}$ rate of convergence.
If a quantity $Y$ is calculated through the measure of the variables $\{y_i\}_{i\leq I}$ with the formula,
\begin{equation}
Y = f(y_1,\ldots,y_I),
\end{equation}
then the variance $\sigma_Y^2$ of $Y$ can be approximated by
\begin{equation}
\label{4_b_variance_estimator_comb}
\sigma_Y^2 = \sum_{i=1}^I \left( \left(\partial_{y_i} f\right)^2 \sigma_{y_i}^2
+ 2 \sum_{j > i} \left(\partial_{y_i} f\right)\left(\partial_{y_j} f\right) \sigma_{y_iy_j}^2  \right),
\end{equation}
where $\sigma_{xy}^2 = \langle xy\rangle - \langle x \rangle \langle y\rangle$.
Using Eq. \eqref{4_real_summation} we can apply this formula to estimate the variance $\sigma_{W'}$ anywhere in phase space, but because of its simple formulation thanks to Eq. \eqref{4_a_neg_zernike}, we will only study it at the origin $(0,0)$:
\begin{equation}
\label{4_b_polynomials_errors}
\sigma^2_{W'}(0,0) = \frac{1}{(J-1)L^2}  \sum_{m=0}^M \left( \sigma_{a_0^m}^2 
+ 2\sum_{k > m}^M (-1)^{m+k} \sigma_{a_0^m a_0^k}^2 \right).
\end{equation}
Notice that in this case the variance estimator formula of Eq. \eqref{4_b_variance_estimator_comb} is not an approximation anymore due to the linear combination nature of Eqs. \eqref{4_real_summation} or \eqref{4_a_neg_zernike}.
We can compute an estimate of $\sigma_{a_0^m}$ when computing the coefficients $w_n^m$ in the same way we did with Eqs. \eqref{4_b_kernel_error_estimator} and \eqref{4_b_kernel_error_estimator_bis}.
Figure \ref{fig_estim_err} shows estimation of the reconstruction errors for different states using Eq. \eqref{4_b_kernelirt_error} and \eqref{4_b_polynomials_errors}.
We have found that the value of $k_c$ has very little influence on $\sigma_{W''}$ at the center of phase space.
On the contrary $M$ has a strong influence on $\sigma_{W'}(0,0)$. 
However, as was shown in Figs.\ref{fig_photon_comp}, \ref{fig_thermal_comp} and \ref{fig_input_comp}, far from the origin the polynomial series tomography algorithm shows less uncertainties.
\begin{figure}[!ht]
\includegraphics[width=0.7\columnwidth]{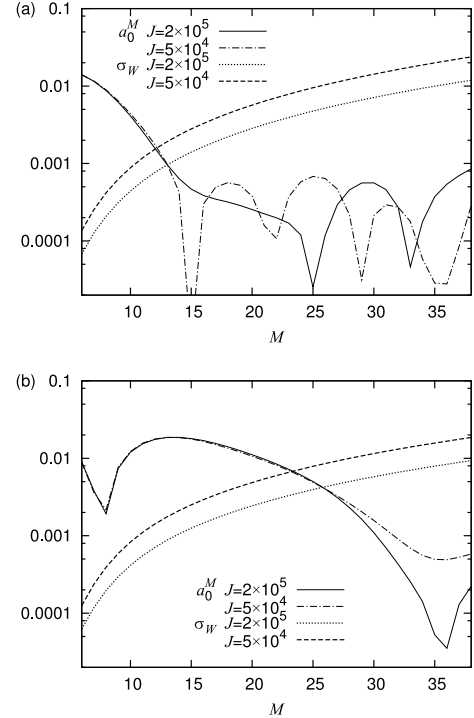}
\caption{\label{fig_neg_convergence} 
Effect of $M$ on the convergence of $W'(0,0)$ and the magnitude of $\sigma_{W'}(0,0)$.
(a) Thermal state with $\langle\hat{n}\rangle=1$, rejection sampling.
(b) Experimental photon subtracted squeezed vacuum state (same data as in Fig.\ref{fig_input_comp}). 
}
\end{figure}

We also assumed the convergence error due to finite truncation $N$ and $M$ of the expansion to be smaller than the statistical error itself.
This can be checked in the algorithm by iteratively calculating $\sigma^2_{W'}(0,0)$ for increasing values of $M$ and stop when the magnitude of the $M^\text{th}$ and last coefficient $w_0^M$ is less than $\sigma^2_{W'}(0,0)$ (see Fig. \ref{fig_neg_convergence}).
This technique can be repeated independently for every point of phase space $(q,p)$, and different values of $N$ and $M$ can even be used for different points of phase space.


\subsection{Monte Carlo error estimation}

\begin{figure}[!ht]
\includegraphics[width=0.9\columnwidth]{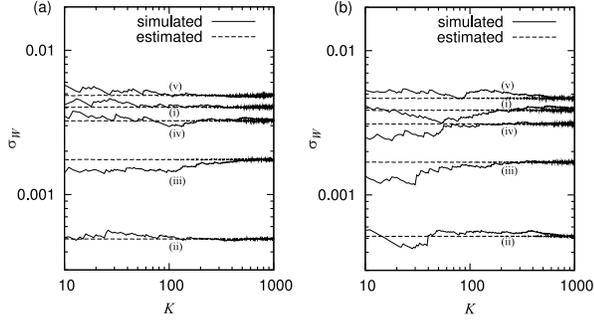}
\caption{\label{fig_simu_err}
Comparison between Monte Carlo simulation and direct estimation of $\sigma_W(0,0)$.
Black curves are the estimation of $\sigma_W(0,0)$ with Monte-Carlo simulation using $K$ data sets.
Dashed curves are the direct estimation of $\sigma_W(0,0)$ using Eqs. \eqref{4_b_kernelirt_error} and \eqref{4_b_polynomials_errors} for the $K^\text{th}$ data set.
(a) Data sets of $J=10^5$ points generated using rejection sampling for the state $0.8 \ket{1}\bra{1} + 0.2 \ket{0}\bra{0}$.
(b) Data sets of $J=10^5$ points generated with bootstrapping resampling from the same experimental data as Fig. \ref{fig_input_comp}.
(i) Filtered back-projection tomography with $k_c = 7$;
(ii) polynomial series tomography with $M=10$;
(iii) $M=20$;
(iv) $M=30$;
(v) $M=40$.
}
\end{figure}

Independently from the estimators of the previous paragraph, we also use Monte Carlo simulations to generate many synthetic data sets and evaluate the reconstruction errors.
This method is easily applied if we know precisely which state $\ketpsi$ is under investigation.
For example, we can choose a known density matrix or Wigner function and calculate the associated marginal distribution $p(x,\theta)$. 
From this marginal distribution we generate $K$ synthetic data sets of $J$ points $\{(x_j,\theta_j)\}_j^{(k)}$ using, for example, rejection sampling. 
With the algorithm of our choice we repeat the tomography reconstruction and calculate a set of $K$ Wigner function $\{W^{(k)}\}_k$.
Finally for a given point of phase space $(x_0,p_0)$, we calculate the average value $\bar{W}_0$ of the set $\{W^{(k)}\}_k$:
\begin{equation}
\bar{W}_0 = \frac{1}{K} \sum_{k=1}^K W^{(k)}(x_0,p_0),
\end{equation}
and obtain an estimate of the error $\sigma_{\bar{W}}$ at point $(x_0,p_0)$ by
\begin{equation}
\sigma^2_{\bar{W}} = \frac{1}{K} \sum_{k=1}^K \left( W^{(k)}(x_0,p_0) - \bar{W}_0 \right)^2. 
\end{equation}
Since it is a Monte Carlo based simulation, every quantity shows again a $1/\sqrt{K}$ convergence rate.

With experimental data, we can sample $p(x,\theta)$ only once and therefore we need a technique to generate the synthetic data sets after the experimental measurement.
Resampling is the easiest approach and here we estimate the reconstruction error of experimental data sets with the bootstrapping resampling method \cite{Chapman94}.
The results of both techniques are illustrated in Fig. \ref{fig_simu_err} and overall there is a good agreement between the estimated values of Monte Carlo simulations and the predicted value of $\sigma_W(0,0)$ using Eq. \eqref{4_b_kernelirt_error} or \eqref{4_b_polynomials_errors}.


\subsection{Distance to a target state}

\begin{figure}[!ht]
\includegraphics[width=0.9\columnwidth]{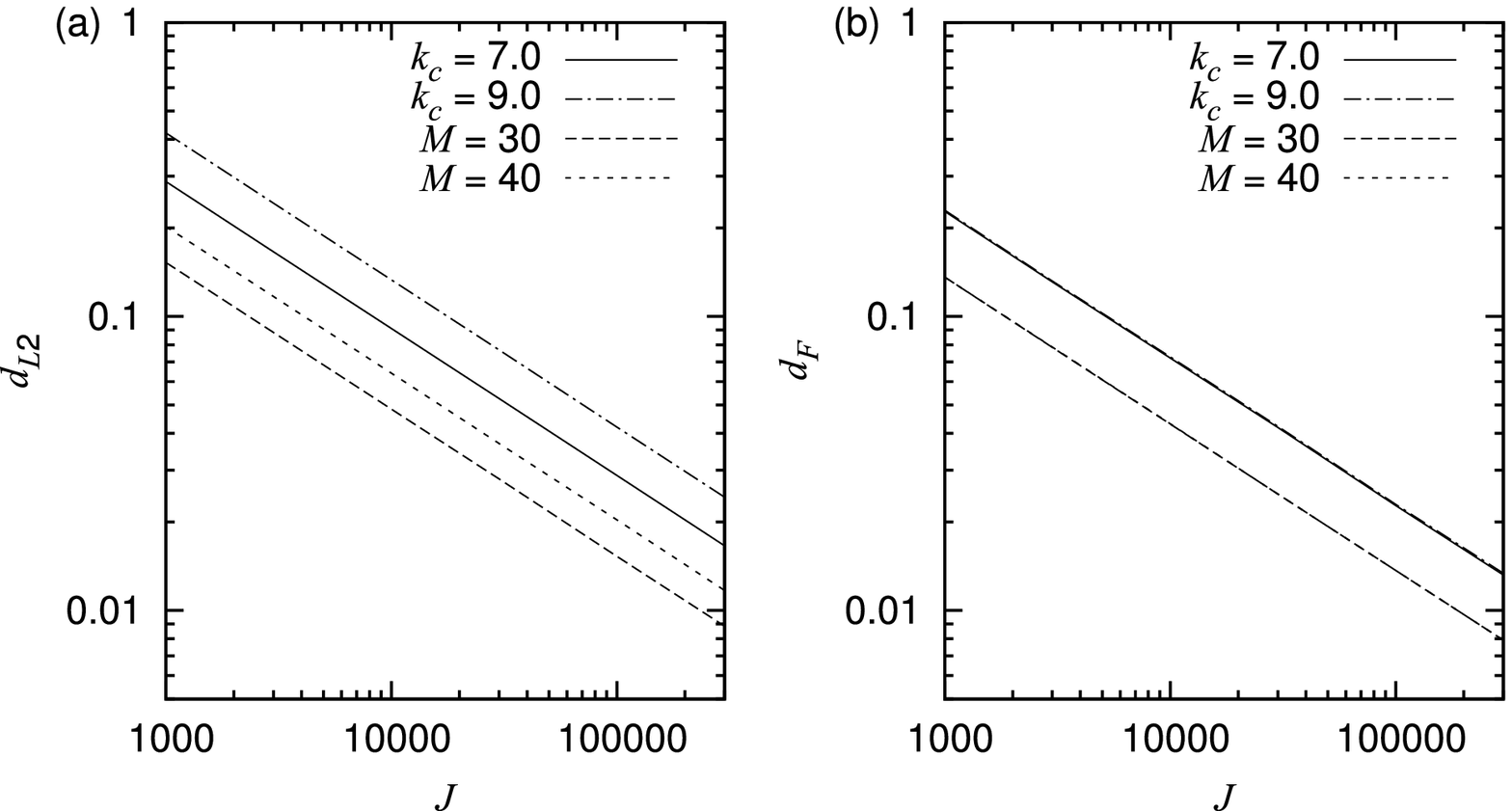}
\caption{\label{fig_distance_thm}
Estimation of the distance between the target thermal state of mean photon number $\langle\hat{n}\rangle = 1$ and reconstructed quantum states averaged over 1000 samples of $J$ data points for different tomography settings.
(a) $L2$ distance $\left\langle d_{L2}(W_\text{target},W_\text{tomo}) \right\rangle$.
(b) Frobenius distance $\left\langle d_{F}(\drho_\text{target}, \drho_\text{tomo}) \right\rangle$.
}
\end{figure}

\begin{figure}[!ht]
\includegraphics[width=0.9\columnwidth]{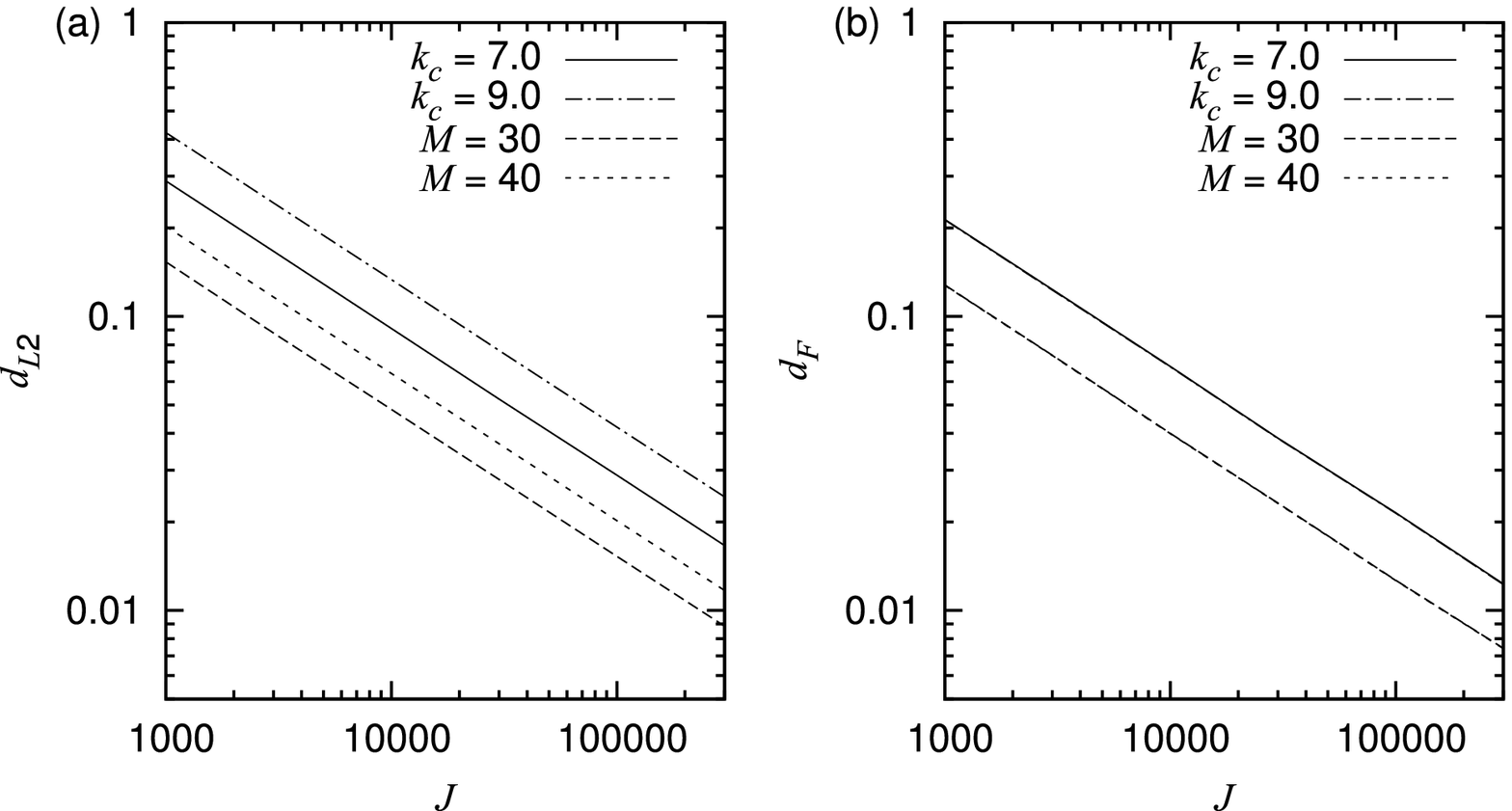}
\caption{\label{fig_distance_moi}
Estimation of the distance between the target state $0.8 \ket{1}\bra{1} + 0.2 \ket{0}\bra{0}$ and reconstructed quantum states averaged over 1000 samples of $J$ data points for different tomography settings.
(a) $L2$ distance $\left\langle d_{L2}(W_\text{target},W_\text{tomo}) \right\rangle$.
(b) Frobenius distance $\left\langle d_{F}(\drho_\text{target}, \drho_\text{tomo}) \right\rangle$.
}
\end{figure}

\begin{figure}[!ht]
\includegraphics[width=0.9\columnwidth]{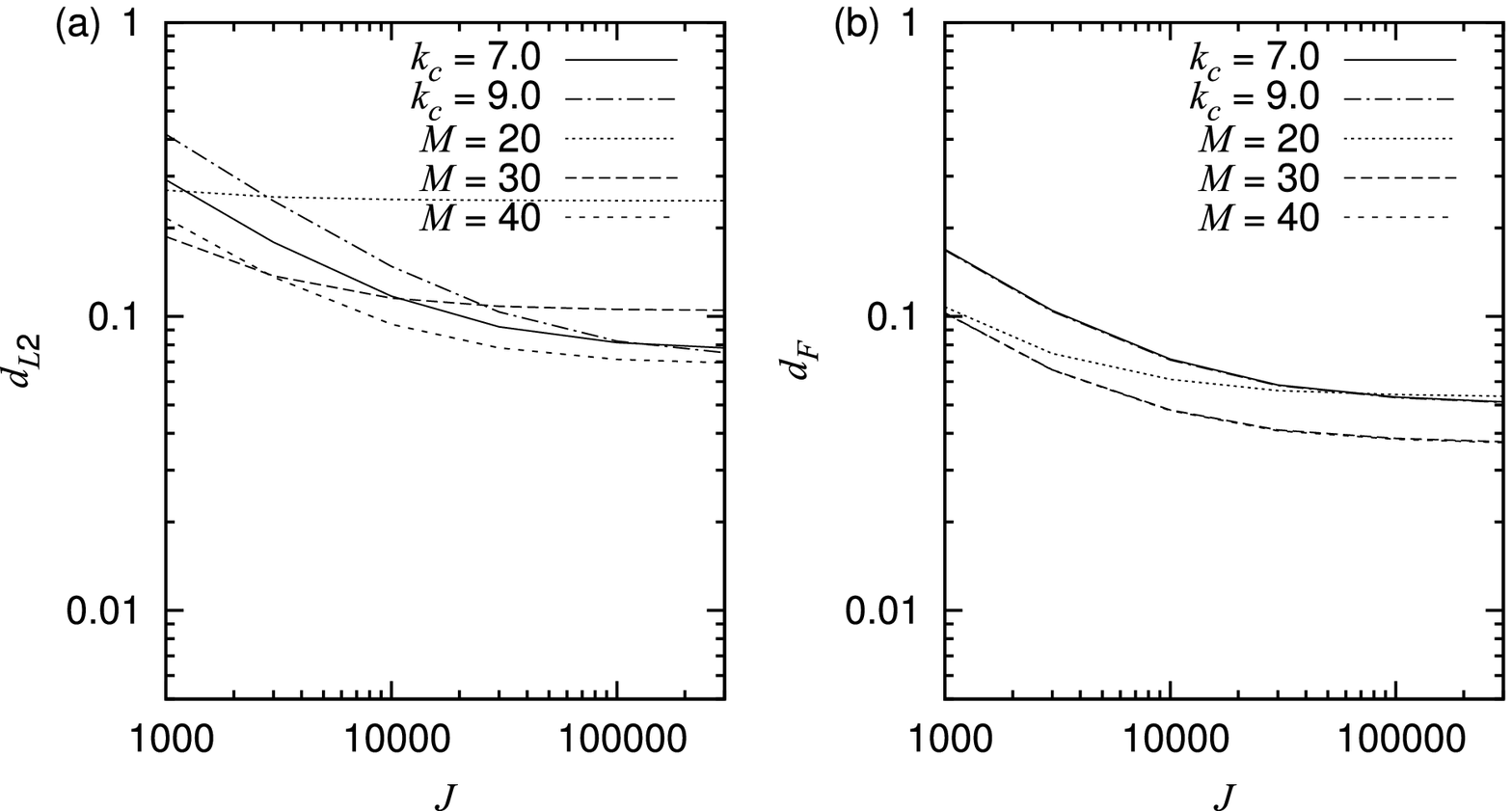}
\caption{\label{fig_distance_cat}
Estimation of the distance between the target odd Schroedinger's cat state $\propto\ket{\alpha} - \ket{-\alpha}$ with $\langle\hat{n}\rangle = 3$ and reconstructed quantum states averaged over 1000 samples of $J$ data points for different tomography settings.
(a) $L2$ distance $\left\langle d_{L2}(W_\text{target},W_\text{tomo}) \right\rangle$.
(b) Frobenius distance $\left\langle d_{F}(\drho_\text{target}, \drho_\text{tomo}) \right\rangle$.
}
\end{figure}

To conclude this comparative study of polynomial series expansion and filtered back-projection-based tomography, we numerically estimate in this final paragraph the distance between some original target quantum state and reconstructed states using both algorithms.
For this purpose we will consider one distance for the Wigner function and one distance for the density matrix.
We use the $L2$ Euclidian distance $d_{L2}(.,.)$ for the Wigner function defined by
\begin{equation}
\label{4_d_dist_l2}
d_{L2}(W_A,W_B) = \left( \int\int dx dp \left|W_A(x,p) - W_B(x,p)\right|^2 \right)^{1/2},
\end{equation}
and with the Frobenius norm $\nm{.}_F$ defined by
\begin{equation}
\nm{A}_F = \sqrt{ \tr\left( A^*A \right) } = \left(\sum_{i,j} |A_{ij}|^2\right)^{1/2},
\end{equation}
we define a distance $d_F(.,.)$ for density matrix as
\begin{equation}
\label{4_d_dist_fr}
d_F(\drho_A,\drho_B) = \nm{\drho_A-\drho_B}.
\end{equation}
First we choose a target state and derive its exact Wigner function $W_\text{target}$ and density matrix $\drho_\text{target}$.
We then evaluate the distances from the target state according to Eqs. \eqref{4_d_dist_l2} and \eqref{4_d_dist_fr} using as before Monte Carlo sampling techniques.
Rather than averaging a reconstructed state over many simulated data sets, we average the distance computed over many reconstructed states and estimate the numbers:
\begin{equation}
\left\langle d_{L2}(W_\text{target},W_\text{tomo}) \right\rangle
 \text{ and }
\left\langle d_{F}(\drho_\text{target}, \drho_\text{tomo}) \right\rangle.
\end{equation}
Numerical simulation results are shown in Figs. \ref{fig_distance_thm}-\ref{fig_distance_cat} for, respectively, a thermal state with $\langle\hat{n}\rangle = 1$, a mixture of vacuum and one-photon state $0.8 \ket{1}\bra{1} + 0.2 \ket{0}\bra{0}$, and an odd Schroedinger's cat state with $\langle\hat{n}\rangle = 3$.
In agreement with the previous results on tomography uncertainties, we observe that polynomial series expansion tomography performs better than filtered back-projection for these two first cases.
In the case of the Schroedinger's cat state $\propto\ket{\alpha}-\ket{-\alpha}$, both distances behave differently for higher $J$ and tend to reach a precision limit which depends on the tomography algorithm and settings.
Although the exact cause of this saturation is unknown, we believe it is due to the significantly more complex structure of the Schroedinger's cat state.
According to our simulations, it seems to depend only on the radial and angular precision settings, more precisely on parameters $M$, $N$, and $k_c$.
In this case again, polynomial series expansion proves to reach a higher precision level than filtered back-projection for a relevant range of tomography settings.
To conclude this paragraph, it is interesting to notice that in the case of the $d_{L2}(.,.)$ distance there is an intrinsic limitation on the precision of polynomial series expansion tomography due to the circular geometry of the reconstruction space \cite{Pawlak02}.
This could be the reason for the saturation phenomenon visible in Fig. \ref{fig_distance_cat}.


\section{Conclusion}

We have shown and demonstrated a technique for optical homodyne tomography based on polynomial series expansion of the Wigner function.
In Sec.II we have given the basis of the usual filtered back-projection algorithm and explained the main reason for its weak performances against statistical noise.
We have also introduced the projection-slice theorem and the relation between phase space, Fourier space and the marginal distribution.
In Sec.III we have shown that it is possible to link three families of orthogonal functions between these three spaces to decompose $p(x,\theta)$ the marginal distribution, $W(q,p)$ the Wigner function, and their Fourier transforms.
We have shown that the Radon transform preserves the orthogonality of these families and therefore takes an especially simple form in this case.
In Sec.IV we have explained and applied to experimental and simulated data the most straightforward implementation of that technique with a direct linear estimation of the coefficients of the polynomial series expansion.
We have also provided estimators of the reconstruction errors and shown that it performs better than filtered back-projection tomography with respect to reconstruction artifacts and statistical errors.
More precisely, polynomial series tomography is superior with fewer experimental data points and when higher radial resolution is needed for higher photon number states.
These results are confirmed when looking at the distance between a chosen target state and states reconstructed with both tomography techniques.
Furthermore this technique exploits the projection slice theorem directly and therefore is faster than convolution based filtered back-projection.
Finally we remark that it is in principle possible to use the maximum likelihood technique to find the set of coefficients $w_m^n$ that maximizes the probability of measuring the experimentally measured data set.


\begin{acknowledgments}
This work was partly supported by the Strategic Information and Communications R\& D Promotion (SCOPE) program of the Ministry of Internal Affairs and Communications of Japan, Project for Developing Innovation Systems, Grants-in-Aid for Scientific Research, Global Center of Excellence, Advanced Photon Science Alliance, and Funding Program for World-Leading Innovative R\&D on Science and Technology (FIRST) commissioned by the Ministry of Education, Culture, Sports, Science and Technology of Japan, and ASCR-JSPS, the Academy of Sciences of the Czech Republic and the Japanese Society for the Promotion of Science.
\end{acknowledgments}


\bibliographystyle{aipproc}

\begin{thebibliography}{99}

\bibitem[Smithey(1993)]{Smithey93}
{D. T. Smithey, M. Beck, M. G. Raymer, A. Faridani, Phys. Rev. Lett. \textbf{70}, 1244 (1993).}

\bibitem[Vogel(1989)]{Vogel89}
{K. Vogel, H. Risken, Phys. Rev. A \textbf{40}, 2847 (1989).}

\bibitem[Radon(1917)]{Radon17}
{J. Radon, Berichte der Sachsischen Akadamie der Wissenschaft \textbf{69}, 262 (1917)[J.Radon (by P. C. Parks), IEEE Transactions on medical imaging MI-5, 170(1986)].}

\bibitem[D'Ariano(1995)]{Dariono95}
{G. M. D'Ariano, U. Leonhardt, H. Paul, Phys. Rev. A \textbf{52}, 1801 (1995).}

\bibitem[Leonhardt(1996)]{Leonhardt96}
{U. Leonhardt, M. G. Raymer, Phys. Rev. Lett. \textbf{76}, 1985 (1996).}

\bibitem[Drobny(2002)]{Drobny02}
{G. Drobny, V. Buzek, Phys. Rev. A \textbf{65}, 053410 (2002).}

\bibitem[Hradil(1997)]{Hradil97}
{Z. Hradil, Phys. Rev. A \textbf{55}, R1561 (1997).}

\bibitem[Lvovsky(2004)]{Lvovsky04}
{A. I. Lvovsky, J. Opt. B: Quantum Semiclass. Opt. \textbf{6}, S55 (2004).}

\bibitem[Moyacessa(19937)]{Moya93}
{H. Moya-Cessa, P. L. Knight, Phys. Rev. A \textbf{48}, 2479 (1993).}

\bibitem[Deleglise(2008)]{Deleglise08}
{S. Del\'eglise, I.Dotsenko, C. Sayrin, J. Bernu, M. Brune, J.-M. Raymond, S. Haroche, Nature \textbf{455}, 510-514 (2008).}

\bibitem[Hawkins(1986)]{Hawkins86}
{W. G. Hawkins, H. H. Barrett, SIAM J. Numer. Anal. \textbf{23}, 873 (1986).}

\bibitem[Rouze(2006)]{Rouze06}
{N. C. Rouze, V. C. Soon, G. D. Hutchins, Pattern Recognition Lett. \textbf{27}, 636-642 (2006).}

\bibitem[Leonhardt(1997)]{Leonhardt}
{U. Leonhardt, \textit{Measuring the Quantum State of Light} (Cambridge University Press, Cambridge, 1997).}

\bibitem[Bracewell(1956)]{Bracewell56}
{R. N. Bracewell, Aust. J. Phys. \textbf{9}, 198 (1956).}

\bibitem[Stark(1981)]{Stark81}
{H. Stark, J. W. Woods, I. Paul, R. Hingorani, IEEE Trans, Biomed. Eng. \textbf{28}, 496-505 (1981).}

\bibitem[Ourjoumtsev(2006)]{Ourjoumtsev06}
{A. Ourjoumtsev, R. Tualle-Brouri, J. Laurat, and P. Grangier, Science \textbf{312}, 83 (2006).}


\bibitem[Born Wofl(1997)]{BornWolf}
{M. Born, E. Wolf, \textit{Principles of Optics} (Cambridge University Press, Cambridge, 1999).}

\bibitem[Deans(2007)]{Deans07}
{S. R. Deans, \textit{The Radon Transform and Some of Its Applications} (Dover Publications, New York, 2007).}

\bibitem[Zeitler(1974)]{Zeitler74}
{E. Zeitler, Optik \textbf{39}, 396-415 (1974).}

\bibitem[Prata(1989)]{Prata89}
{A. Prata, W. V. T. Rusch, Appl. Opt. \textbf{28}, 749-754 (1989).}

\bibitem[Lee and al(2010)]{Lee11}
{N. Lee, H. Benichi, Y. Takeno, S. Takeda, J. Webb, E. Huntington, and A. Furusawa, Science \textbf{332}, 330 (2011).}

\bibitem[Chapman(1994)]{Chapman94}
{B. Efron, R. J. Tibshirani, \textit{An Introduction to the Bootstrap} (Chapman \& Hall/CRC, New York, 1994).}


\bibitem[Pawlak(2002)]{Pawlak02}
{M. Pawlak, S. X. Liao, IEEE Trans. Info. Theory \textbf{48}, 2736-2753 (2002).}


\end{thebibliography}


\end{document}